\newif\ifieee  
\ieeetrue
\newif\ifieeefinal
\ieeefinaltrue
\newif\iftodo       
\todofalse
\newif\iftodoshort  
\todoshortfalse
\ifieee
  \ifieeefinal
    \documentclass{IEEEtran}
  \else
    \documentclass[a4paper,onecolumn,draftcls]{IEEEtran}
  \fi 
\else
  \documentclass[11pt,fleqn]{article}
\fi 
\ifieee
\else
\fi
\iftodo
  \usepackage[german,english]{babel}   
  \usepackage{color}
\fi
\usepackage{amsmath}
\usepackage{amssymb}
\ifieee
  \usepackage{theorem}
\else
  \usepackage{amsthm}
\fi
\usepackage[dvips]{graphicx}
\usepackage{xspace}
\newcommand{\myline}{\hfill\\} 
\newcommand{\emptypage}%
{
  \newpage
  \vspace*{10cm}
  \pagebreak
}
\ifieee 
  \newcommand{\textin}{\hangindent0.5cm\hangafter1\myline} 
   
\else 
  \newcommand{\textin}{\hangindent0.75cm\hangafter1\myline} 
   
\fi
\ifieee
  \theoremstyle{plain}
  \theoremheaderfont{\bf}
  {\theorembodyfont{\it}
  \newtheorem{theorem}{Theorem}[section] 
  \newtheorem{proposition}[theorem]{Proposition}
  \newtheorem{lemma}[theorem]{Lemma}
  \newtheorem{corollary}[theorem]{Corollary}
  \newtheorem{conclusion}[theorem]{Conclusion}
  }
  {\theorembodyfont{\rm}
  \newtheorem{definition}[theorem]{Definition}
  }
  {\theorembodyfont{\rm}
  \newtheorem{remark}[theorem]{Remark}
  \newtheorem{note}[theorem]{Note}
  \newtheorem{cit}[theorem]{Citation}
  }
\else
  \theoremstyle{plain}
  \newtheorem{theorem}{Theorem}[section]  
  \newtheorem*{theorem*}{Theorem}
  \newtheorem{proposition}[theorem]{Proposition}
  \newtheorem*{proposition*}{Proposition}
  \newtheorem{lemma}[theorem]{Lemma}
  \newtheorem*{lemma*}{Lemma}
  \newtheorem{corollary}[theorem]{Corollary}
  \newtheorem*{corollary*}{Corollary}
  \newtheorem{conclusion}[theorem]{Conclusion}
  \newtheorem*{conclusion*}{Conclusion}
  \theoremstyle{definition}
  
  \newtheorem*{definition*}{Definition}
  \theoremstyle{remark}
  \newtheorem{remark}[theorem]{Remark}
  \newtheorem*{remark*}{Remark}
  
  \newtheorem*{note*}{Note}
  
  \newtheorem*{cit*}{Citation}

\fi 
\ifieee\else
  
\fi
\newcommand{\mathcomma}{\:,\:\:}
\newcommand{\mapset}[2]{#1\longrightarrow #2}
\newcommand{\mapto}[2]{#1\longmapsto #2}
\usepackage{dsfont}
\newcommand{\CC}{\ensuremath{\mathds{C}}}   
\newcommand{\RR}{\ensuremath{\mathds{R}}}

\newcommand{\NN}{\ensuremath{\mathds{N}}}

\newcommand{\mdef}{:=}
\newcommand{\mfed}{=:}
\newcommand{\oneover}[1]{\frac{1}{#1}}

\renewcommand{\d}{\partial} 
\newcommand{\norm}[1]{\lVert#1\rVert}

\newcommand{\abs}[1]{\lvert#1\rvert}

\newcommand{\generate}[1]{\left\langle #1 \right\rangle}

\newcommand{\scalprod}[1]{<\!\!#1\!\!>}

\newcommand{\fnorm}[1]{\lVert#1\rVert_{\scriptscriptstyle\text{F}}}
\newcommand{\bigfnorm}[1]{\left\lVert#1\right\rVert_{\scriptscriptstyle\text{F}}}

\newcommand{\Unity}{\ensuremath{\mathbf{1}}\xspace}
\newcommand{\Zero}{\ensuremath{\mathbf{0}}\xspace}

\newcommand{\obar}[1]{\smash[t]{\overset{\rule[-0.5pt]{0.75ex}{0.25pt}}{#1}}}
\newcommand{\ubar}[1]{\smash[b]{\underset{\rule[5pt]{0.75ex}{0.25pt}}{#1}}}

\newcommand{\leftsup}[1]{{}^{\scriptscriptstyle #1}}

\newcommand{\textmatrix}[1]{
   \left(\begin{smallmatrix}#1\end{smallmatrix}\right)}
\newcommand{\eqmatrix}[1]{\begin{pmatrix}#1\end{pmatrix}}
\newcommand{\smallperp}{\scriptscriptstyle\perp}

\newcommand{\follows}{\ensuremath{\quad\Longrightarrow\quad}}

\newcommand{\equivalentshort}{\ensuremath{\;\Longleftrightarrow\;}}
\newcommand{\tends}{\ensuremath{\longrightarrow}}
\newcommand{\tendshort}{\ensuremath{\rightarrow}}

\DeclareMathOperator{\tr}{tr}
\renewcommand{\div}{\ensuremath{\text{div}}\,}
\DeclareMathOperator{\grad}{grad}

\DeclareMathOperator{\id}{id}

\DeclareMathOperator{\diag}{diag}
\DeclareMathOperator{\res}{Res}

\DeclareMathOperator{\dimr}{\dim_{\RR}}

\newcommand{\E}{\ensuremath{\mathsf{E}}}
\newcommand{\n}{\ensuremath{T}\xspace}
\renewcommand{\k}{\ensuremath{n_t}}
\newcommand{\kk}{\ensuremath{n_r}}
\newcommand{\Gc}[1]{\ensuremath{G^{\CC}_{#1}}}  
\newcommand{\Vc}[1]{\ensuremath{V^{\CC}_{#1}}}  
\newcommand{\gkn}{\Gc{\k,\n}\xspace}
\newcommand{\vkn}{\Vc{\k,\n}\xspace}
\newcommand{\kbein}{\ensuremath{\textmatrix{\Unity\\\Zero}}}
\usepackage{bbm}
\newcommand{\menge}[1]{\mathbbm{#1}}
\newcommand{\ordered}[2]{\binom{\menge{#1}}{#2}}

\newcommand{\noverk}{\frac{\n}{\k}}
\newcommand{\tnoverk}{\tfrac{\n}{\k}}
\newcommand{\Un}{\ensuremath{U(\n)}\xspace}
\newcommand{\Uk}{\ensuremath{U(\k)}\xspace}

\newcommand{\herm}{\leftsup{\mathcal{H}}\!}
\newcommand{\daggered}[1]{\ensuremath{#1^{\dagger}#1}}
\newcommand{\deltakc}{\ensuremath{\obar{\Delta}}}
\newcommand{\ddeltakc}{\ensuremath{\deltakc^{\dagger}\deltakc}}
\newcommand{\deltauc}{\ensuremath{\underline{\Delta}}}
\newcommand{\ddeltauc}{\ensuremath{\deltauc^{\dagger}\deltauc}}
\renewcommand{\d}{\ensuremath{d}}
\newcommand{\dkc}{\ensuremath{\obar{d}}\xspace}
\newcommand{\duc}{\ensuremath{\ubar{d}}\xspace}
\newcommand{\ddist}{\ensuremath{\dkc\mathrm{dist}}\xspace}
\newcommand{\sigmakc}{\ensuremath{{\obar{\sigma}}}}
\newcommand{\sigmauc}{\ensuremath{{\ubar{\sigma}}}}
\newcommand{\varrhokc}{\ensuremath{\obar{\varrho}}}
\newcommand{\varrhouc}{\ensuremath{\ubar{\varrho}}}
\renewcommand{\div}{\ensuremath{\mathcal{D}iv}}
\newcommand{\divkc}{\ensuremath{\overline{\mathcal{D}iv}}}
\newcommand{\divuc}{\ensuremath{\underline{\mathcal{D}iv}}}
\newcommand{\s}{\ensuremath{s}}
\newcommand{\skc}{\ensuremath{\obar{s}}}
\newcommand{\suc}{\ensuremath{\ubar{s}}}
\newcommand{\p}{\ensuremath{p}}
\newcommand{\pkc}{\ensuremath{\obar{p}}\xspace}
\newcommand{\puc}{\ensuremath{\ubar{p}}\xspace}
\newcommand{\pdist}{\ensuremath{\pkc\mathrm{dist}}\xspace}
\newcommand{\sdist}{\ensuremath{\skc\mathrm{dist}}\xspace}

\newcommand{\sym}{\ensuremath{\mathrm{sym}}}
\title{Geometrical relations between space time block code designs and
  complexity reduction}
\author{Oliver Henkel (henkel@hhi.fraunhofer.de)\\
        Fraunhofer German-Sino Lab for Mobile Communications -- MCI\\
        Einsteinufer 37, 10587 Berlin, Germany
\thanks{Journal reference: IEEE Trans. Inform. Theory {\bf 52}, no 12 (2006),
5324--5335.
Personal use of this material is permitted. However, permission to
reprint/republish this material for advertising or promotional purposes or for
creating new collective works for resale or redistribution to servers or lists,
or to reuse any copyrighted component of this work in other works must be
obtained from the IEEE}
}
\begin{document}
\maketitle
\begin{abstract}
In this work, the geometric relation between space time block code design
for the coherent channel and its non-coherent counterpart is exploited to
get an analogue of the information theoretic inequality 
$I(X;S)\le I((X,H);S)$ in terms of diversity. It provides a lower bound on
the performance of non-coherent codes when used in coherent scenarios.
This leads in turn to a code design decomposition result splitting coherent
code design into two complexity reduced sub tasks. 
Moreover a geometrical criterion for high performance space time code
design is derived. 
\end{abstract}
%
%
%
%
%
%
%
%
%
%
%
%
%
%
\section{Introduction}
In MIMO (Multiple Input Multiple Output) systems space time coding schemes
have been proven to be an appropriate tool to exploit the spatial diversity 
gains. Two distinct scenarios are common, whether the channel coefficients
are known (coherent scenario) \cite{tar.ses.cal},
to the receiver or not (non-coherent scenario) \cite{hoc.mar.2}.
Prominent coherent codes are the well known Alamouti scheme \cite{ala} and
general orthogonal designs \cite{tar.jaf.cal}. A more flexible coding
scheme are the so-called linear dispersion codes. They have been introduced in
\cite{has.hoc} and were further investigated in \cite{goh.dav}. A full rate
high performing example is the recently discovered Golden code
\cite{bel.rek.vit}. 
Genuine non-coherent codes have been proposed in \cite{tar}, but most of
the research efforts in the literature focus on differential schemes,
introduced in \cite{hoc.swe}, since differential codes usually provide
higher data rates than comparable non differential codes. High performing
examples have been constructed in \cite{sho.has.hoc.swe},
\cite{han.ros},\cite{lia.xia},\cite{wan.wan.xia}. 
However, in both (coherent and non-coherent) cases most research effort has
been undertaken for space 
time block codes with quadratic 2-by-2, resp. \k-by-\k\ code
matrices (\k\ denotes the number of transmit antennas). Although linear
dispersion codes are not restricted to quadratic 
shape of the design matrices the block length is not a free design
parameter when the number of transmit antennas is held fixed (compare the
asymptotic guidelines in \cite{goh.dav}).

In general \cite{hoc.mar.1} the signal matrices are of 
rectangular shape of size $\n\times\k$ with unitary columns. The
corresponding coding spaces for the coherent and 
non-coherent scenario are the complex Stiefel and Grassmann manifolds
respectively. Typically the number $\k$ of transmit antennas is a small
number due to hardware limitations, while the block 
length $\n$ can be chosen rather large, upper bounded only by the coherence
length of the channel. Inspired from \cite{bar.nog}, 
in \cite{hen-transinf1} a general analysis of packings in the Stiefel and Grassmann
manifold revealed, that the achievable squared minimal distance (i.e. the
squared diameter of the decision regions for decoding) grows
proportionally with the block length $\n$, more precisely the
following proposition holds \cite{hen-transinf1}):
\begin{proposition}\label{prop.dmin-lowerbound}\textin
For any $\n \ge 2\k$ set $D_{\k,\n}=\k(2\n-\k)$ (coherent channel), resp.
$D_{\k,\n}=2\k(\n-\k)$ (non-coherent channel). Then for any prescribed rate
$R=\oneover{\n}\log\abs{\mathcal{C}}$ there exist space time block codes
$\mathcal{C}$ with rate $R$ and minimal distance $\d_0$ satisfying
\begin{equation}
   \d_0
     \ge C \sqrt{\rho\noverk}
         \left( \oneover{2} \right)^{\frac{\n R}{D_{\k,\n}}}
    \quad\mathcomma\text{(provided $\rho\ge1$)}
\end{equation}
for some constant $C>0$ depending on the channel knowledge at the
receiver. Since the rightmost term is monotonically increasing as a
function of \n, the receiver performance increases proportionally to $\noverk$. 
\end{proposition}
Having the common literature (see above) in mind, this result
comes rather unexpected and further research effort seems
promising. However, explicit code constructions have already been achieved
in \cite{hen-globecom06}. Moreover, the Proposition
\ref{prop.dmin-lowerbound} becomes even more
important when considering  space frequency code design: The schemes 
\cite{boe.bor.pau}, \cite{boe.bor} indicate, that the relevant coding spaces
are certain subsets of (large dimensional) Stiefel and Grassmann
manifolds. Thus considering these coding spaces in general may
be of considerable importance for space frequency code designs. Explicit
space frequency constructions can be found in e.g. \cite{hen-isit05}.

In the present work it will be shown, how general 
space time block code designs can be decomposed into two 'smaller' pieces
with reduced design complexity 
(Theorem \ref{thm.code-diversity-compose}), both
already in the focus of current research.
The achieved result can be seen as complementary to that of Kammoun and Belfiore
\cite{kam.bel}, who presented a coding scheme for non-coherent channel
space time block codes in terms of coherent channel ones, compare Remark
\ref{rem.tower} for further implications.

The key observation is the quite intuitive but technically not obvious diversity
monotonicity (Proposition \ref{prop.code-embedding}), which states that the
performance of each 
non-coherent channel space time block code grows when considered as a coherent
channel code. This turns out to be due to some higher resolution of the
coherent channel receiver, reflecting the information theoretic relation
between the system designs. 

Further insight on the performance is obtained by local analysis of
diversity, leading to the overall picture of space time coding as a
constrained sphere packing problem. It reveals additional structures not
obvious from the traditional point of view, proposing high performance
design criteria 
(Conclusion \ref{conc.cg-points}, \ref{conc.cg-points-in-v}) 
and adding a further estimate 
(Proposition \ref{prop.explicit-embedding-bounds}) 
to the diversity embedding.
By the way all results are obtained in the spirit of geometrical methods in
space time coding theory. 

The remainder of this work is organized as follows. Section \ref{s.models}
introduces the 
basic models for the channels and coding spaces (with emphasis on their
geometrical structures), fixes notation and
conventions used throughout this work. Section
\ref{s.performance-embedding} defines diversities for the coherent/non-coherent
channel cases as our fundamental performance measure and analyses their
interrelations, culminating in the 
embedding and decomposition results mentioned above.
Section \ref{s.extremal-diversity} focuses on the local analysis of
diversity and the connection to the sphere packing problem, exploring its
consequences. 
Finally, the main results will be summarized for
concise reference together with remaining open questions.
\section{Channel model and coding spaces}\label{s.models}
In this section the basic channel model will be presented, leading to the
Stiefel and Grassmann manifolds as coding spaces. These spaces will be
introduced with emphasis on their topological metric structures induced by
the maximum likelihood receivers. The geometric relation between the coding
spaces is precisely expressed by the principal fiber structure, which is
also introduced here. Although the geometric terms used in this work will
be defined (as far as it seems necessary to understand the concepts), the
reader who prefers rigorous definitions is 
invited to consult standard text books e.g. \cite{boothby,conlon} (manifolds),
\cite{conlon,gal.hul.laf,poor} (homogeneous spaces, Lie groups), and/or
\cite{conlon,poor} (principal fibers). For the particular case of the
(complex) Stiefel and Grassmann manifolds an introduction to their real
counterparts aimed at non-specialists is \cite{ede.ari.smi}.
\subsection{Channel model}
We consider the Rayleigh flat fading MIMO (multiple input multiple output)
channel without channel knowledge 
at the transmitter and maximum likelihood decoding at the receiver as
described in \cite{hoc.mar.1} (with normed expected power 
$\sum_j\E\abs{s_{ij}}^2=1$ per time step, $i=1,\dots,\n$, $\E$ denotes
expectation):
\begin{equation}\begin{gathered}
   X= \sqrt{\rho}\, S H + W
   \mathcomma\\
   S=(s_{ij}) \in \CC^{{\n} \times {\k}},\quad
   H \in \CC^{{\k} \times {\kk}},\quad
   X,W \in \CC^{{\n} \times {\kk}}
\end{gathered}\end{equation}
whereas ${\n}$ denotes the coherence time of the channel (respectively the
block length of the signals), ${\k},{\kk}$ denote the number of transmit,
resp. receiver antennas, $W \sim \text{i.i.d. } {\cal CN}(0,1)$ is the
noise, $H \sim \text{i.i.d. }{\cal CN}(0,1)$ the channel matrix and $S,X$
denote the transmitted, resp. received signal with SNR (signal to noise
ration) $\rho$.
The (ergodic) channel capacity is defined by the supremum of the
mutual information 
\begin{equation} 
   C = \frac{1}{\n} \sup_{p(S)} I((X,H);S)
   \mathcomma\text{resp. }
   C = \frac{1}{\n} \sup_{p(S)} I(X;S)
\end{equation} 
for the coherent (resp. non-coherent) channel, 
and we define the rate $R$ of the code ${\cal C}$ by
\begin{equation}
   R := \frac{1}{\n}\log\abs{{\cal C}}
\end{equation}
The normalization by $1/\n$ is merely a convention to have the block length
$\n$ as a free design parameter of the code, such that codes with distinct
block length are comparable.
\subsection{Coding spaces}
Hochwald and Marzetta \cite[Theorem 1 and 2]{hoc.mar.1} have shown, that
signals $S$ of the form $S = \oneover{\sqrt{\k}} \Phi {\cal E}$
are optimal with respect to the channel capacity
(due to the central limit theorem tending to $C_0$ defined below, when
$\n\tends\infty$), if the receiver does not know the channel. 
More precisely one has ${\n} \ge {\k}$, 
${\cal E}=\diag(\epsilon_i) \in \CC^{{\k} \times {\k}}$ with $\epsilon_i$
non-negative, 
$\E \epsilon_i^2 = \n$ stochastic independent from $\Phi$, obeying
$\Phi^{\dagger}\Phi=\Unity_{\k}$ ($\k\times\k$-unit matrix), $\Phi$
therefore being canonically an element of the 
complex Stiefel manifold \vkn\ defined below. 
In \cite{hoc.mar.2} Hochwald/Marzetta, and more generally Zheng/Tse in
\cite[Lemma 8]{zhe.tse} have shown that the optimal 
energy allocation ${\cal E}$ of the antennas equals (asymptotically in $\rho$) 
${\cal E}=\sqrt{{\n}}\Unity$, thus 
\begin{equation}
   S=\sqrt{\noverk}\,\Phi
\end{equation} 
for ${\k} \ge {\kk}$, ${\n} \ge 2{\kk}$. The
signal $S=\sqrt{\tnoverk}\,\Phi$ then carries the total energy
$\norm{S}_F^2={\n}$, thus the transmitter sends with unit power per time
step. In this case the mutual information
(ergodic in the channel realizations) $I(X;S)$ depends only on the
subspace in $\CC^{\n}$ spanned by the columns of $\Phi$,
not on $\Phi$ itself \cite{zhe.tse}. This is reflected by the fact,
that scalings and linear combinations of the columns of $\Phi$ are indistinguishable
for the detector, when the channel is non-coherent. Therefore these transformations
cannot carry any information and we end up with signals
$S \in \sqrt{\tnoverk}\,\gkn$, \gkn denoting the complex Grassmann manifold
of $\k$-dimensional linear subspaces of $\CC^{\n}$.

For the coherent channel the capacity has been calculated by
Telatar \cite{tel} to
\begin{equation}
   C_0 := \E \log\det\left( \mathbf{1}+\frac{\rho}{{\k}}H^{\dagger}H
                     \right)
\end{equation} 
Assuming the same energy allocation ${\cal E}=\sqrt{\n}\Unity$ one can
justify, that now the asymptotically optimal signal space consists of
signals $S \in \sqrt{\tnoverk}\,\vkn$. 

We focus on both signal designs
in this article, sometimes called unitary space time modulation in the
literature (introduced in \cite{hoc.mar.2}). 
\subsection{Coherent channel: The Stiefel manifold ${\vkn}$}
The (complex) Stiefel manifold defined by
\begin{equation}\label{e.defvkn} 
   \vkn \mdef \{ \Phi\in\CC^{\n \times \k} \,|\, \Phi^{\dagger}\Phi=\Unity \}
\end{equation}
is diffeomorphic to a coset space with respect to the unitary group
$U(\n)$ of \n-by-\n unitary matrices:
\begin{equation}
   \vkn \cong
      \Un \left/ \textmatrix{
            \Unity & \Zero\\
            \Zero & U(\n-\k)
          }\right.\mathcomma
   \Phi \cong (\Phi,\Phi^{\smallperp})\kbein
\end{equation}
whereas $\cong$ means 'diffeomorphic to'. From this equivalence we obtain 
\begin{equation}\begin{split}
   \obar{D}_{\k,\n} &\mdef\dimr\vkn=\dimr\Un-\dimr U(\n-\k)\\
                    &=\k(2\n-\k)
\end{split}\end{equation} 
for free. Since the elements of the Stiefel manifolds are \k-dimensional
orthonormal bases, they are called \emph{\k-frames}.
Geometrically the coset representation of \vkn is interpreted as a
so-called \emph{homogeneous space} 
\begin{equation}
   \left(\Un \overset{\pi^U_V}{\tends} \vkn;\, U(\n-\k)\right)
\end{equation}
This means that each $\Phi\in\vkn$ is the image of a projection $\pi^U_V$ from some unitary
\n-by-\n matrix $U$ (in the coset representation
$\pi^U_V$ is simply the projection on the first \k\ columns of $U=(\Phi,\Phi^{\smallperp})$) 
and for each $\Phi,\Psi\in\vkn$ there exist an unitary $U_{\Phi\Psi}$ with
$\Psi=U_{\Phi\Psi}\Phi$. The latter property is obviously fulfilled and
called a \emph{transitive left action} of the 
group \Un on \vkn (the defining property for \vkn being a homogeneous
space), while the former property means that $\pi^U_V$ is invariant 
with respect to the \emph{right action} of $U(\n-\k)$ on \vkn.

As a linear algebraic convention used in this work, eigenvalues $\lambda_i$
and singular values $\sigma_i$ of matrices will be arranged in decreasing
order, thus $\lambda_1\ge\dots\ge\lambda_{\k}$, and
$\sigma_1\ge\dots\ge\sigma_{\k}$.  

A code ${\cal C}^V$ for the coherent channel model is given by a discrete set 
${\cal C}^V = \{\Phi_i\}\subset\vkn$. 
At the receiver the maximum likelihood decision reads (see \cite{hoc.mar.2})
\begin{equation}\label{e.mlkc}
   \Phi_{\text{ML}} = \arg\min_{\forall_{\Phi\in{\cal C}}}
   \bigfnorm{X-\sqrt{\rho\tnoverk}\,\Phi H}
\end{equation}
whereas $X=\sqrt{\rho\noverk}\,\Psi H+W$ is the received signal. Throwing
away the noise term allows a formulation of a code design criterion in the
signal space \vkn, induced from the ML receiver: The maximization of the
pairwise distances \dkc, given by
\begin{gather}
   \label{e.dkc}
   \dkc(\Phi,\Psi) 
       \mdef \fnorm{\deltakc} 
       = \sqrt{\sum_{i=1}^{\k} \sigma_i^2(\deltakc)}
       = \norm{\sigmakc}
\end{gather}
where we have set 
\begin{gather}
   \deltakc:=\Phi-\Psi \\
   \label{e.sigmakc}
   \sigmakc:=(\sigmakc_1,\dots,\sigmakc_{\k})
   \mathcomma
   \sigmakc_i:=\sigma_i(\deltakc)\in [0,2]
\end{gather}
Thus coding corresponds to a packing problem on the metric space
$\big(\vkn,\dkc\big)$\footnote{We will see in section
  \ref{s.performance-embedding}, that this is only an approximation of the
  design criterion, but the importance of the packing gain will become
  clear in section \ref{s.extremal-diversity}}. 
Note, that by
\begin{equation}\label{e.sigmakc-invariance} 
   \sigma_i(U(\Phi-\Psi)v)=\sigma_i(\Phi-\Psi)
   \mathcomma \forall_{U\in\Un,\,v\in\Uk}
\end{equation}
the metric \dkc remains invariant under left or right multiplication
of its arguments with unitary matrices (also denoted as 
\emph{left invariance} resp. \emph{right invariance}):
\begin{equation}\label{e.dkc-invariance}
   \dkc(U\Phi v,U\Psi v) = \dkc(\Phi,\Psi)
\end{equation}
This property is one motivation for the geometric picture of the Stiefel
manifold as a homogeneous space with its corresponding left and right
actions. Furthermore for each singular value $\sigmakc_i$ holds
\begin{equation}\label{e.sigmalambda}\begin{split}
   \tfrac{1}{2}\sigmakc_i^2
    & = \tfrac{1}{2}\sigma_i^2(\deltakc) 
      = \tfrac{1}{2}\lambda_i(\ddeltakc)\\
    & = \lambda_i(\Unity-\herm(\Phi^{\dagger}\Psi))
      = 1-\lambda_{\k-i+1}(\herm(\Phi^{\dagger}\Psi))
\end{split}\end{equation}
whereas $\herm{M}\mdef 1/2(M+M^{\dagger})$ denotes the hermitian part.
\subsection{Non-Coherent channel: The Grassmann manifold ${\gkn}$}
The (complex) Grassmann manifold \gkn is the set of all $\k$-dimensional
(complex) linear subspaces of $\CC^{\n}$:
\begin{equation}
   \gkn \mdef \{ \generate{\Phi} \,|\, \Phi\in\vkn \}
\end{equation}
whereas $\generate{\Phi}$ denotes the column space of $\Phi$. 
Since $\mapto{\Phi}{\generate{\Phi}}$ is a projection invariant under all
\k-by-\k\ unitary basis transformations we get the coset representation
\begin{equation}
   \gkn \cong \Un \left/ \textmatrix{
                           \Uk   & \Zero\\
                           \Zero & U(\n-\k)
                         }\right.\mathcomma
   \generate{\Phi} \cong \Phi{\Phi^1}^{-1}
\end{equation}
$(\Phi^1 \mdef (\Unity,\Zero)\Phi)$ and 
\begin{equation}\begin{split}
   \ubar{D}_{\k,\n} 
     &\mdef \dimr\gkn\\
     &= \dimr\Un-\dimr U(\k)-\dimr U(\n-\k)\\
     &=2\k(\n-\k)
\end{split}\end{equation}
Note that the coordinate representation 
$\generate{\Phi} \cong \Phi{\Phi^1}^{-1}$ holds only locally in general
(since it requires $\Phi^1$ to have full rank),
but it turns out, that this representation covers all but a set of measure
zero and we abandon this distinction between local and global properties in
the sequel and drop the distinction between \gkn and its coordinate
domain.

Again we have a geometrical reformulation in terms of the homogeneous space
\begin{equation}
   \left(\Un \overset{\pi^U_G}{\tends} \gkn;\, U(\k) \times U(\n-\k)\right)
\end{equation}
whereas the transitive left action now reads 
$\generate{\Psi}
 =U_{\generate{\Phi}\generate{\Psi}}\generate{\Phi}
 \mdef\generate{U_{\generate{\Phi}\generate{\Psi}}\Phi}$
(e.g. choose $U_{\generate{\Phi}\generate{\Psi}}=U_{\Phi\Psi}$). 
The projection $\pi^U_G$ is now invariant with respect to the combined
right action of $ U(\k) \times U(\n-\k)$, because not only the orthogonal
complement of the columns in $\Phi$ has been neglected, but also the
particular choice of the spanning \k-frame: Each 
$(\Phi,\Phi^{\smallperp})\textmatrix{u & \Zero \\ \Zero & v}$ 
represents the same
space $\generate{\Phi}$ for arbitrary $u\in U(\k)$, $v\in U(\n-\k)$.

To simplify matters let us assume $\k\le\n/2$ whenever we are in contact
with the Grassmann manifold. This is 
no restriction, since for $\k\ge\n/2$ we can always switch to the orthogonal
complement of the subspaces under consideration. 
Given now two elements $\generate{\Phi},\generate{\Psi}\in\gkn$ then there
exist $\k$ principal angles
$0\le\vartheta_1\le\dots\le\vartheta_{\k}\le\pi/2$ between  
$\generate{\Phi}$ and $\generate{\Psi}$. They are defined
successively by the critical values
$\arccos\abs{\scalprod{v_i, w_i}}$, $i=1,\dots,\k$ (in increasing order),
of $\mapto{(v,w)}{\arccos\abs{\scalprod{v, w}}}$
where the unit vectors $v,w$ vary over 
$\{v_1,\dots,v_{i-1}\}^{\perp} \subset \generate{\Phi}$, 
respectively
$\{w_1,\dots,w_{i-1}\}^{\perp} \subset \generate{\Psi}$, compare \cite{bjo.gol}.
The components of the vector of principal angles 
$\vartheta\mdef(\vartheta_1,\dots,\vartheta_{\k})$ 
can be computed by the formula (any representing $\k$-frame will do) \cite{bjo.gol}
\begin{equation}\label{e.principalangles}
   \cos\vartheta_i = \sigma_i(\Phi^{\dagger}\Psi)
\end{equation}
An important application of principal angles on some given pair 
$\generate{\Phi},\generate{\Psi}$ with principal angles 
$\vartheta$ is, that due to the transitivity of the 
unitary group action there exist an unitary $U$, such that $\Psi$ (say) can
always be translated into  
$\kbein=U\Psi$ and in $U\generate{\Phi}=\generate{U\Phi}$ one can choose a
basis such that we end up with the canonical representing $\k$-frames
\begin{equation}\label{e.standard0-vpair}
   \Psi_0 = \eqmatrix{\Unity\\\Zero} \mathcomma
   \Phi_0 = \eqmatrix{(\cos\vartheta_i)\\(\sin\vartheta_i)\\\Zero}
\end{equation}
(where 
 $(\cos\vartheta_i)\mdef\diag(\cos\vartheta_i)_{i=1,\dots,\k}\in\RR^{\k\times\k}$) 
for the translated spaces 
$\generate{\Psi_0}=U\generate{\Psi}$, 
$\generate{\Phi_0}=U\generate{\Phi}$.
Note, that the demand to \emph{choose} the appropriate basis $\Phi_0$ in 
$U\generate{\Phi}$ is mandatory, in general there is no
$U\in\Un$ which translates the $\k$-frames $\Psi$, $\Phi$ simultaneously
into $\Psi_0$, $\Phi_0$. 
\subsection{The principal fiber structure ${P_G^V}$}
The natural relationship between the homogeneous spaces
\vkn and \gkn is subsumed in the canonical \emph{principal fiber bundle} structure
\begin{equation}\label{e.vgpfb}
   P^V_G \mdef \big( \vkn \overset{\pi^V_G}{\tends} \gkn;\, U(\k) \big)
\end{equation}
which (locally) embeds \gkn into \vkn by choosing a representing $\k$-frame
$\Phi$ which spans the subspace $\generate{\Phi}$. However there remains
the freedom of multiplication with arbitrary unitary 
matrices $u\in\Uk$ from the right (all of them have the same image under
the projection $\pi^V_G$), and for practical applications it is
necessary to specify a unique
choice for $\Phi$ and $u$, given $\generate{\Phi}$ (simultaneously for all
$\generate{\Phi}\in\gkn$, not only for pairs as in
\eqref{e.standard0-vpair}). But locally this can always be achieved and we
do not want to go into details here. 
The term 'principal fiber bundle' means a generalization of the term
'homogeneous space', where now the \emph{total space} \vkn no longer need
to be a group and the \emph{base space} \gkn is a projection $\pi^V_G$ of
the total space which is invariant under a right action of $U(\k)$. The set of
all elements $\Phi u$ is called a \emph{fiber} over $\generate{\Phi}$.

This geometrical point of view makes clear, that we can consider 
codes ${\cal C}^G\subset\gkn$ for the non-coherent channel as discrete subsets
of \vkn in virtue of the local embedding of \gkn into \vkn. But one
motivation for the introduction of all these perhaps unfamiliar geometrical
terms is to clarify the relationship between the coding spaces, i.e. that
there is no canonical representation of 
${\cal C}^G$ in \vkn. In practical applications this peculiarity is often
overlooked, since common mathematics software packages already use
certain conventions when representing subspaces in terms of singular value
decompositions. Furthermore we will see that the unitary left and right
actions on the coding spaces lead naturally to the diversity embedding results
derived in the next section. These results are geometrical in nature rather
than linear algebraic, but only in the geometric context it becomes clear,
that they are not obvious at all, since they relate distinct metric
structures. For the Stiefel manifold the relevant metric structure has
already been defined in (\ref{e.dkc}) and for the Grassmann manifold we
will define it next.

We consider codes ${\cal C}^G \subset \gkn$ 
always as discrete subsets of $\vkn$
and the maximum likelihood criterion for the non-coherent channel receiver reads
now (\cite{hoc.mar.2}) 
\begin{equation}\label{e.mluc}
   \Phi_{\text{ML}} = \arg\max_{\forall_{\Phi\in{\cal C}}}
   \bigfnorm{\sqrt{\rho\tnoverk}\,\Phi^{\dagger}X}
\end{equation}
whereas $X=\sqrt{\rho \noverk}\,\Psi H+W$ is the received signal. To obtain
a design criterion in the signal space \gkn we throw away the noise
term (as in the coherent channel case) and pass from $\Psi H$ to 
$\Psi\in\gkn\subset\vkn$ (this operation does not change the column space
of $\Psi$). Setting 
\begin{gather}
   \deltauc := \Phi^{\dagger}\Psi \\
   \label{e.fnorm.uc}
   \fnorm{\deltauc}^2 
     = \sum_{i=1}^{\k} \sigma_i^2(\deltauc)
     \overset{\eqref{e.principalangles}}{=} \sum_i \cos^2\vartheta_i
   = {\k} - \sum_i \sin^2\vartheta_i
\end{gather}
(note that \eqref{e.fnorm.uc} does not depend on the choice of the
representing $\k$-frame, thus represents really an entity on \gkn), and 
\begin{equation}\label{e.sigmauc}
   \sigmauc=(\sigmauc_1,\dots,\sigmauc_{\k})
   \mathcomma
   \sigmauc_i:=\sigma_i(\deltauc)=\cos\vartheta_i 
\end{equation}
the ML criterion demands the maximization of the pairwise distances
\begin{equation}\label{e.duc}\begin{split}
   \duc(\Phi,\Psi) 
   & \mdef \sqrt{\k-\fnorm{\deltauc}^2}
     = \sqrt{\tr\bigl(\Unity-\daggered{\deltauc}\bigr)}\\
   & = \sqrt{\sum_{i=1}^{\k} \bigl(1-\sigmauc_i^2\bigr)} 
     = \sqrt{\sum_{i=1}^{\k} \sin^2\vartheta_i}
\end{split}\end{equation}
Formally $\duc$ is defined on all of \vkn, but independent of the choice of
the representing $\k$-frame as already indicated. Of course, it is a metric
in the strict sense only as a function on \gkn (known as the 'chordal'
distance, compare \cite{con.har.slo,bar.nog}), turning again the coding
problem into a packing problem in $\big(\gkn,\duc\big)$. It shares the
invariance properties of the coherent channel metric \dkc, but
satisfies even more: 
\begin{equation}\label{e.duc-invariance}
   \duc(U\Phi v,U\Psi w)=\duc(\Phi,\Psi)
\end{equation}
by
\begin{equation}\label{e.sigmauc-invariance}
   \sigma_i((U\Phi v)^{\dagger}(U\Psi w))=\sigma_i(\Phi^{\dagger}\Psi)
   \mathcomma
   \forall_{U\in\Un,\,v,w\in\Uk}
\end{equation}
\section{Performance analysis: Diversity}\label{s.performance-embedding}
In practical settings, where $\rho,\n \ll \infty$, the receiver metrics
$\duc,\dkc$ fail to be the sole code design criteria. 
Denoting the pairwise error probability of mistaking one symbol for another
at the receiver generically as $P_{ij}$ one gets the union upper bound 
\begin{equation}
   \oneover{\abs{{\cal C}}}\sum_i\sum_{j\neq i}P_{ij}
\end{equation}
for the exact error probability.
This section deals with the pairwise error probability Chernov
bound, more precisely with the \emph{diversity}, which is essentially the reciprocal
of the Chernov bound. It turns out, that the receiver metric coincides 
with the first order term of the diversity
and the highest order term leads to the so called diversity
product (further analyzed in section \ref{s.extremal-diversity}).
Adopting the diversity as the major performance measure, section
\ref{ss.embedding} investigates the connection between non-coherent and coherent
channel designs and its consequences for code design.
For convenience we fix the pair $(\Phi,\Psi)$ of code
symbols throughout this section and suppress their notation as
function arguments. 

For the coherent channel case the pairwise error probability has been calculated in
\cite{hoc.mar.2} to
\begin{equation}\label{e.pep-kc}\begin{split}
  & P(\Phi,\Psi) = 
    \sum_{\{\obar{\alpha}_j\}} \res_{\omega=\imath\obar{\alpha}_j} \\
  &    \left\{
         -\frac{1}{\omega+\imath/2}
            \prod_{\substack{i=1\\\sigmakc_i>0}}^{\k}
               \left[
                  \frac{1}
                       {\rho\noverk\,\sigmakc_i^2(\omega^2+\obar{\alpha}_i^2)}
               \right]^{{\kk}}
      \right\}
\end{split}\end{equation}
with
$\obar{\alpha}_i := \sqrt{\frac{1}{4}+\frac{1}{(\rho\n/\k)\sigmakc_i^2}}$. 
Analogously for the non-coherent channel case holds \cite{hoc.mar.2} 
\begin{equation}\begin{split}
 & P(\Phi,\Psi) = 
    \sum_{\{\ubar{\alpha}_j\}} \res_{\omega=\imath\ubar{\alpha}_j}\\
 &  \left\{
     -\frac{1}{\omega+\imath/2}\prod_{\substack{i=1\\\sigmauc_i<1}}^{\k}
     \left[
       \frac{1+(\rho\noverk)}
       {(\rho\noverk)^2(1-\sigmauc_i^2)
         (\omega^2+\ubar{\alpha}_i^2)}
     \right]^{{\kk}}
   \right\}
\end{split}\end{equation}
with 
$\ubar{\alpha}_i := \sqrt{\frac{1}{4}
                          +\frac{1+(\rho {\n}/{\k})}
                                {(\rho {\n}/{\k})^2(1-\sigmauc_i^2)}}$.

For both cases the we have the Chernov bound
\begin{equation}\label{e.chernov}
   P \le \frac{1}{2} \left( \prod_{i=1}^{\k}
               \left[
                 1+ \varrho\, \sigma_i^2
               \right]\right)^{-{\kk}}
\end{equation}
whereas (coherent channel)
\begin{gather}
   \label{e.varrhokc}
   \varrho = \varrhokc 
      \mdef \frac{\rho\n}{4\k} \\
   \sigma_i = \sigmakc_i 
\end{gather}
respectively (non-coherent channel)
\begin{gather}
   \label{e.varrhouc}
   \varrho 
      = \varrhouc 
      \mdef \frac{\varrhokc^2}{\varrhokc+1/4}
          = \frac{(\rho\noverk)^2}{4(1+\rho\noverk)} \\
   \sigma_i = \sqrt{1-\sigmauc_i^2}
\end{gather}

The term in parentheses in \eqref{e.chernov} is called (pairwise)
diversity 
\begin{equation}
   \div \mdef \prod_{i=1}^{\k}
               \left[
                 1+ \varrho\, \sigma_i^2
               \right]
\end{equation}
and we take it as our basic performance measure for codes. Rewriting \div\ 
as a polynomial in $\varrho$ requires the use of elementary symmetric
polynomials defined by  
$\sym^{\k}_0 \mdef 1$,
$\sym^{\k}_j(x_1,\dots,x_{\k}) 
 = \sum_{I_j\in\ordered{\k}{j}} x^{}_{I_j}
 = \sum_{I_j\in\ordered{\k}{j}} x_{i_1}\cdots x_{i_j}$
(with $\ordered{\k}{j}\mdef
 \{(i_1,\dots,i_j)\in\NN^j\,|\, 1\le i_1<\dots< i_j \le \k\}$),
$j=1,\dots,\k$. With the abbreviation
\begin{equation}
   s_j
     \mdef \sym^{\k}_j(\sigma_1^2,\dots,\sigma_{\k}^2)
     = \sum_{I_j\in\ordered{\k}{j}} s^{2}_{I_j}
\end{equation}
we find generically
\begin{equation}\label{e.diversity} 
  \div = \sum_{i=0}^{\k} s_i\varrho^i
\end{equation}
The first and highest order coefficient of this polynomial are of
particular importance, since they dominate the diversity in the low and
high SNR regime respectively. They are called \emph{diversity sum} and
\emph{diversity product} respectively, and are given by
\begin{gather}
   \d = \sqrt{\s_1} = \norm{\sigma} \\
   \label{e.p}
   \p \mdef \sqrt{\s_{\k}} = \sigma_1\cdots\sigma_{\k}
\end{gather}
The diversity sum is our familiar metric 
$\dkc=\norm{\sigmakc}=\fnorm{\deltakc}$ \eqref{e.dkc}, resp.
$\duc=\norm{\sigmauc}=\sqrt{\tr(\Unity-\ddeltauc)}=\norm{\sin\vartheta}$ \eqref{e.duc}.
The diversity product acts as a regularity criterion for the positive
semidefinite matrix $\ddeltakc$, resp. 
$\Unity-\ddeltauc$: In the coherent channel case 
$\pkc^2=\det\ddeltakc>0$ is known as diversity criteria (resp. rank
criteria or determinant criteria) in the literature
(e.g. \cite{tar.ses.cal}). In the non-coherent channel case 
$\puc^2=\det(\Unity-\ddeltauc)
      =(1-\sigmauc^2_1)\dots(1-\sigmauc^2_{\k})
      \overset{\eqref{e.sigmauc}}{=}\sin^2\vartheta_1\dots\sin^2\vartheta_{\k}>0$
measures the positivity of the principal angles between
$\generate{\Phi}$ and $\generate{\Psi}$. 

All terms $\skc_{I_j}$, resp. $\suc_{I_j}$ in the diversity
expansion possess the invariance properties induced by
(\ref{e.sigmakc-invariance}), resp. (\ref{e.sigmauc-invariance}). 
Therefore the analysis in this section applies to all terms in
\eqref{e.diversity} and the result can be stated in closed form for the
full diversity, rather than only to its first and highest order coefficient. 

Specializing \eqref{e.diversity} to the non-coherent channel case, one checks
easily that the coefficients are formally defined on all of \vkn, but
independent of the choice of the representing $\k$-frame. 
Note that the coherent and non-coherent channel diversities are formally similar
due to \eqref{e.diversity}, but the constituting singular values 
\eqref{e.sigmakc}, \eqref{e.sigmauc} reflect the underlying topological
structures induced by the maximum likelihood receivers \eqref{e.mlkc},
\eqref{e.mluc} (resp. the metrics \dkc, \duc). And these structures are
entirely distinct.   
\subsection{Embedding properties}\label{ss.embedding}
Now let us investigate the relation between the non-coherent and coherent channel
diversity quantities. From the information theoretic inequality
$I(X;S)\le I((X,H);S)$ between the corresponding mutual informations we
expect such a relation satisfied by the diversity. 
The ranges for $\sigmakc$ \eqref{e.sigmakc} and $\sigmauc$
\eqref{e.sigmauc} indicate, that the coherent channel receiver may benefit
from some higher 'resolution', but if and how this carries over to the
diversity is not obvious and requires a
rigorous proof. The investigations of this section give an affirmative
answer to that conjecture.

By a slight abuse of notation let us define the 'fiber minima' 
of $\skc^{}_{I_j}$ with respect to the fibers of $P_G^V$ \eqref{e.vgpfb}
as
\begin{equation}
   \sdist^{}_{I_j}(\Phi,\Psi) \mdef
      \min_{\substack{\Phi\in{\pi^V_G}^{-1}(\generate{\Phi})\\
                      \Psi\in{\pi^V_G}^{-1}(\generate{\Psi})}}
            \skc^{}_{I_j}(\Phi,\Psi)\\
\end{equation}
Then we obtain 
\begin{lemma}\textin
   Let $\generate{\Phi},\generate{\Psi}\in\gkn$ separated by principal
   angles $\vartheta_1,\dots,\vartheta_{\k}$. 
   Then $\forall_{I_j\in\ordered{\k}{j}}$
   \begin{equation}
      \sdist^{}_{I_j}(\Phi,\Psi)
         = \sqrt{2^{j}\prod_{i\in I_j}(1-\cos\vartheta_{\k-i+1})}
   \end{equation}
   holds.
\end{lemma}
\vspace{-3ex}\myline
\begin{proof}\myline
   Due to left invariance of $\skc^{}_{I_j}$ we can 
   switch to the canonical $\k$-frame bases $\Psi_0$, $\Phi_0$
   \eqref{e.standard0-vpair} of $U\generate{\Psi}$, $U\generate{\Phi}$. 
   With $\Phi_0(u)\mdef\Phi_0 u$ ($u\in\Uk$), running through the fiber over
   $\generate{\Phi_0}$, $\deltakc(u)\mdef\Phi_0(u)-\Psi_0$, and
   $\tfrac{1}{2}\deltakc(u)^{\dagger}\deltakc(u)
      =\Unity-\herm((\cos\vartheta_l)u)$ 
   (recall, that $(\cos\vartheta_l)=\diag(\cos\vartheta_l)_{l=1\dots\k}$)
   we have
   \begin{equation}\begin{split}
         \skc^2_{I_j}(\Phi_0(u),\Psi_0)
         & = \prod_{i\in I_j} \sigma_i^2(\deltakc(u))\\
         & \overset{\eqref{e.sigmalambda}}{=}
             \prod_{i\in I_j} 2 
             \big\{1-\lambda_{\k-i+1}[\herm((\cos\vartheta_l)u)]\big\}\\
         & \overset{(*)}{\ge} 2^j \prod_{i\in I_j} 
                 \big\{ 1-\sigma_{\k-i+1}
                 \left((\cos\vartheta_l)u\right) \big\}\\
         & = 2^j \prod_{i\in I_j} 
                 \big\{ 1-\sigma_{\k-i+1}
                 ((\cos\vartheta_l)) \big\}\\
         & = 2^j \prod_{i\in I_j} ( 1- \cos\vartheta_{\k-i+1} ) 
   \end{split}\end{equation}
   where $(*)$ comes from the general inequality
   \begin{equation}\label{e.bha-estimate}
      \lambda_i\big(\herm(A)\big) \le \sigma_i(A)
   \end{equation}
   devoted to Fan-Hoffman in \cite[Prop. III.5.1]{bha}. $u=\Unity$
   achieves equality in $(*)$ and this completes the proof.
\end{proof}
In particular we have the fiber distance
\begin{equation}\label{e.dist}\begin{split}
   \ddist(\Phi,\Psi) 
    & \mdef
        \min_{\substack{\Phi\in{\pi^V_G}^{-1}(\generate{\Phi})\\
                      \Psi\in{\pi^V_G}^{-1}(\generate{\Psi})}}
            \dkc(\Phi,\Psi)\\
    & = \sqrt{
          \sum_{I_1\in\binom{\menge{\k}}{1}}\sdist^2_{I_1}
             }
      = \sqrt{2\sum_{i=1}^{\k}(1-\cos\vartheta_i)}
\end{split}\end{equation}
and its analogon for the diversity product
\begin{equation}\label{e.pdist}\begin{split}
   \pdist(\Phi,\Psi) 
    & \mdef
      \min_{\substack{\Phi\in{\pi^V_G}^{-1}(\generate{\Phi})\\
                      \Psi\in{\pi^V_G}^{-1}(\generate{\Psi})}}
            \pkc(\Phi,\Psi)\\
    &  = \sdist^{}_{I_{\k}}
       = \sqrt{2^{\k}\prod_{i=1}^{\k}(1-\cos\vartheta_i)}
\end{split}\end{equation}
We observe, that the fiber minima $\sdist^{}_{I_j}$ for each given pair
$(\Phi,\Psi)$ are realized by the same choice $(\Phi_0,\Psi_0)$, which
justifies the definition
\begin{equation}
   \sdist_j(\Phi,\Psi) \mdef
      \sum_{I_j\in\binom{\menge{\k}}{j}}\sdist^2_{I_j}(\Phi,\Psi)
\end{equation}
(thus we have in particular
$\ddist=\sqrt{\sdist_1}$ and $\pdist=\sqrt{\sdist_{\k}}$)
leading to
\begin{corollary}\label{cor.s-embedding}\textin
   For any pair $\generate{\Phi},\generate{\Psi}\in\gkn$
   we have $\forall_{j=1,\dots,\k}$
   \begin{equation}\label{e.spair.embedding}
      \suc_j(\Phi,\Psi) 
        \le \sdist_j(\Phi, \Psi)
        \le \skc_j(\Phi,\Psi)
   \end{equation}   
\end{corollary}
\vspace{-3ex}\myline
\begin{proof}\myline
   The second inequality holds by definition of \sdist. So let us turn to
   the first inequality and denote the principal angles between
   $\generate{\Phi}$ and $\generate{\Psi}$ by
   $\vartheta=(\vartheta_1,\dots,\vartheta_{\k})$,
   $0\le\vartheta_i\le\frac{\pi}{2}$. We have
   $\suc^2_{I_j}=\prod_{i\in I_j} \sin^2\vartheta_i
                =\prod_{i\in I_j} (1-\cos^2\vartheta_i)$ 
   and
   $\sdist^2_{I_j}=2^j\prod_{i\in I_j}(1-\cos\vartheta_{\k-i+1})
                  =\prod_{i\in I_j} 2(1-\cos\vartheta_{\k-i+1})$.
   For any $0\le\vartheta_i\le\frac{\pi}{2}$
   \begin{equation}
      2\cos\vartheta_i-\cos^2\vartheta_i \le 1
   \end{equation}
   holds, thus
   \begin{equation}\label{e.sij-le-sdistijstar}
      \suc^{}_{I_j}\le\sdist^{}_{I^*_j}
   \end{equation}
   with $I^*_j\mdef(\k-i_1+1,\dots,\k-i_j+1)$ 
   for each $I_j=(i_1,\dots,i_j)$, and by
   \begin{equation}
      \suc_j
       =\sum_{I_j} \suc^2_{I_j}
       \overset{\eqref{e.sij-le-sdistijstar}}{\le}\sum_{I_j} \sdist^2_{I^*_j}
       =\sum_{I_j} \sdist^2_{I_j}
       =\sdist_j
    \end{equation}
    the claim follows.
\end{proof}
Given a function 
$s:\mapset{{\cal C} \times {\cal C}}{\RR}$, let us define 
$s^{\min} \mdef \min_{{\cal C} \times {\cal C}} s$.
Then we state 
\begin{corollary}\label{cor.min-s-embedding}\textin
   \begin{equation}\label{e.min-s-embedding}
      \suc^{\min}_j\le\sdist^{\min}_j\le\skc^{\min}_j
      \mathcomma\quad
      \forall_{j=1,\dots,\k}
   \end{equation}\vspace{-4ex}
\end{corollary}
(unfortunately neither there seems to be a canonical way to determine the
pairs of points, which realize the minima, nor whether this could be
achieved simultaneously for each of the quantities above by a single pair of
points) 
\\[1ex]
\begin{proof}\myline
   Corollary \ref{cor.min-s-embedding} is an easy consequence of
   Corollary \ref{cor.s-embedding}. 
   For each inequality the proof goes the same, so let us take two
   functions $\mathrm{f} \le \mathrm{F}$, $\mapset{X \times X}{\RR}$, $X$ a
   discrete set, for brevity. Then there are two cases 
   \begin{enumerate}
   \item 
      $\mathrm{f}^{\min}=\mathrm{f}(x_0,y_0)
       \le \mathrm{F}(x_0,y_0)=\mathrm{F}^{\min}$ 
      and there is nothing to do
   \item 
      $\mathrm{f}^{\min}=\mathrm{f}(x_0,y_0)$, but
      $\mathrm{F}(x_0,y_0)>\mathrm{F}(x_1,y_1)=\mathrm{F}^{\min}$.
      But then still
      $\mathrm{f}^{\min}\le \mathrm{f}(x_1,y_1)
       \le \mathrm{F}(x_1,y_1)=\mathrm{F}^{\min}$
      holds.
   \end{enumerate}
\end{proof}
On the metric level (the diversity sum) this inequalities provide a
distance gain due to the channel knowledge. 
It increases the resolution of the detector and 
allows the receiver to separate points better 
than the non-coherent channel receiver could do, $\duc\le\dkc$, or equivalently
the unit spheres with respect to 
$\dkc$ occupy smaller volume than the corresponding (embedded) unit spheres
with respect to $\duc$, thus one can pack more $\dkc$-spheres into \vkn
than $\duc$-spheres. 
But due to the famous estimate \eqref{e.bha-estimate} we have proven a
considerable stronger result not confined to the diversity sum, but rather
to any coefficient in the diversity expansion \eqref{e.diversity}. Thus we
are able to relate the inequalities derived so far to the diversity as a
whole: 
Comparing the 'effective' SNRs $\varrhokc$ \eqref{e.varrhokc}, $\varrhouc$
\eqref{e.varrhouc} in the diversity \eqref{e.diversity} demands one
additional estimate (provided $\n\ge\k$)
\begin{equation}\label{e.varrhouc-estimate}
     \frac{4}{\rho}\ubar{\varrho}=\frac{\rho(\tnoverk)^2}{1+\rho\tnoverk}
     \le\noverk
     \follows
     \ubar{\varrho}\le\obar{\varrho}
\end{equation}
thus we have
\begin{proposition}\label{prop.code-embedding}\textin
   For any pair $\generate{\Phi}, \generate{\Psi} \in \gkn$
   \begin{equation}
       \divuc(\Phi,\Psi) \le \divkc(\Phi,\Psi)
   \end{equation}
   holds.
\end{proposition}
\vspace{-3ex}\myline
\begin{proof}\myline
   The proposition follows directly from \eqref{e.spair.embedding}
   and \eqref{e.varrhouc-estimate}
\end{proof}
So we conclude, that the coherent channel maximum likelihood receiver applied to 
${\cal C}^G$ has at least the diversity as the non-coherent channel receiver,
the diversity grows. This approves the
information theoretic inequality $I(X;S)\le I((X,H);S)$ motivating our
analysis.

Having explored the relationship of the embedding $\gkn\subset\vkn$ let us
come to a somewhat complementary scenario, 
which offers the possibility of coding complexity reduction:
Consider a single fiber over $\generate{\Phi}$. Then,
by $\Phi w=(\Phi,\Phi^{\smallperp})\textmatrix{w\\\Zero}$, there
holds a special kind of 'vertical' left invariance, namely
\begin{equation}
   \skc^{}_{I_j}(\Phi u,\Phi v) = \skc^{U}_{I_j}(u,v)
   \mathcomma\quad
   \forall_{I_j\in\ordered{\k}{j}}
   \mathcomma\forall_{u,v \in U(\k)}
\end{equation}
where the right hand side is evaluated in $\Uk=\Vc{\k,\k}$.
Analogously we define for the special case $\n=\k$:
$\varrhokc^U\mdef\frac{\rho}{4}$,
$\skc^U_j\mdef\sum_{I_j\in\ordered{\k}{j}}(\skc^U_{I_j})^2$,
$\divkc^U\mdef\sum_i \skc^{U}_i(\varrhokc^{U})^i$ and we arrive at
\begin{theorem}\label{thm.code-diversity-compose}\textin
   Given codes 
   ${\cal C}^G\subset \gkn \subset \vkn$
   and 
   ${\cal C}^U\subset \Uk$, 
   then the composed code
   ${\cal C}^V\subset \vkn$ given by
   \begin{equation}
      {\cal C}^V \mdef {\cal C}^G \cdot {\cal C}^U 
                  = \left\{\Phi u \,|\, 
                         \Phi\in {\cal C}^G, u\in {\cal C}^U\right\}
   \end{equation}
   satisfies
   \begin{equation}
     \skc^{\min}_j \ge \min\{\sdist^{\min}_j,\skc^{U\min}_j\}
     \mathcomma\quad
     \forall_{j=1,\dots,\k}
   \end{equation}
   and 
   \begin{equation}
      \divkc^{\min} \ge \min\{\divuc^{\min},\smash{\widetilde{\divkc}}^{U \min}\}
   \end{equation}
   holds, whereas
   $\smash{\widetilde{\divkc}}^U\mdef \sum_i (\noverk)^i \skc^U_i(\varrhokc^U)^i$
   (thus the power constraint factor $\sqrt{\noverk}$ sharpens the estimate).
\end{theorem}
\begin{proof}
The theorem follows directly from Corollary \ref{cor.min-s-embedding},
Proposition \ref{prop.code-embedding} and the definition of $\sdist$ as a
fiber minimum.
\end{proof}
Therefore the code design splits up into two parts:
Codes ${\cal C}^G$ represent the familiar coding problem for the non-coherent
channel corresponding to $\gkn$, which has smaller
dimension as the general problem in $\vkn$. 
The code ${\cal C}^U$ represents a coding problem for the coherent channel
in $\Uk=\Vc{\k,\k}$, contributing the dimensions left by 
$\vkn\cong\gkn\times\Uk$ locally.
So both parts represent a somewhat smaller coding problem with respect to
the dimension of the signal spaces. Moreover for both parts the code design
is easier to solve than in $(\vkn,\dkc)$: In $\Uk$ there are
many solutions (i.e. codes) in the common literature, e.g. the Alamouti
scheme for $\k=2$, orthogonal designs for $\k\ge 2$, quasi-orthogonal space
time block codes, and many more. The Grassmannian part is also simpler (not
only concerning dimensions but also) in
structure, because the 'chordal' design metric $\duc$ is 
geometrically more natural than the Euclidean distance measure $\dkc$ (in
terms of their relation to the natural geodesic distance
\cite{hen-transinf1}), thus geometric methods may apply. Also
packings in $(\gkn,\duc)$ are already in the focus of current research,
e.g. \cite{con.har.slo,bar.nog}, whereas \cite{con.har.slo} also
contains explicit constructions for packings in the (real) Grassmann
manifold. In \cite{hen.wun-itg05} a differential geometric connection
(based on \cite{hen-transinf1}) has been developed to construct space time
(and space frequency) codes for for the coherent and non-coherent channel
case. Further research \cite{hen-globecom06} led to space time codes
with reduced design complexity by utilizing
\ref{thm.code-diversity-compose}. 
\begin{remark}\textin
   A related question arises, when one considers the task of given a code
   ${\cal C}^U$, does there exist a code 
   ${\cal C}^V$ 
   with the same rate but better performance than ${\cal C}^U$?
   Concerning the diversity sum $\d$ a partial answer gives \cite{hen-transinf1}:
   The transmit power constraint sets the requirement 
   $\sqrt{\frac{\n}{\k}}\dkc^{\min}\ge\dkc^{U\min}$.
   Since there exist a monotonically increasing lower bound for
   $\dkc^{\min}$ when $\frac{\n}{\k}$ grows (Proposition
   \ref{prop.dmin-lowerbound}) this requirement can be certainly
   fulfilled. This again emphasizes the need for 
   coding strategies in the general coding spaces $\vkn$, $\gkn$, $\n$
   larger than 
   $\k$. However, it remains an open question, whether we can achieve the
   goal by composed codes of the form 
   ${\cal C}^V = {\cal C}^G \cdot {\cal C}^U$.
\end{remark}
\begin{remark}\label{rem.tower}\textin
   A conceptual simple (but computational complex) embedding of \gkn into
   \vkn is given by the parametrization of \gkn with (so-called
   'horizontal') tangents 
   $X^H=\textmatrix{\Zero & -B^{\dagger}\\ B & \Zero}$, 
   $B\in\CC^{(\n-\k) \times \k}$ in
   its total space \Un. In a recent article \cite{kam.bel} it has been
   shown, that coding for the non-coherent channel is under certain 
   assumptions equivalent to coding on 
   the horizontal tangent space, with respect to the coherent channel
   diversity for $\Vc{\k,\n-\k}$. Combining that with Theorem
   \ref{thm.code-diversity-compose} 
   we can roughly state this correspondence
   as $\Vc{\k,\n-\k} \subset \gkn \subset \vkn$,
   which gives rise to a sequence
   $\dots\tendshort \mathcal{C}_{i}^V 
         \tendshort \mathcal{C}_{i+1}^G 
         \tendshort \mathcal{C}_{i+1}^V
         \tendshort\dots$
   of codes with increasing block length $i\cdot\k$, $i=1,2,\dots$
\end{remark}
\section{Extremal properties of the diversity}\label{s.extremal-diversity}
\renewcommand{\d}{\partial}
In this section we examine the distribution of pairwise angles in 
${\cal C}^G$ to find criteria for maximum diversity in particular for the 
combined code ${\cal C}^V={\cal C}^G \cdot {\cal C}^U$ in
$\sqrt{\noverk}\vkn$. We focus on the diversity sum and diversity product,
representing the most important diversity quantities (since they dominate
the small and high SNR regime of diversity) while still being simple
functions of the principal angles.

To get some first insight into the interplay between diversity sum and
product (with respect to a fixed pair $\Phi,\Psi$ of code symbols) we
exploit the homogeneity of the elementary symmetric 
polynomials. For both coherent and non-coherent channel case it is quite natural to write
$\hat{\sigma}\mdef\frac{\sigma}{\norm{\sigma}}$, 
$\hat{\s}_i\mdef\frac{\s_i}{\norm{\sigma}^{2i}}
 =\sym_i(\hat{\sigma}^2_1,\dots,\hat{\sigma}^2_{\k})$,
The importance of this factorization arises from the identity 
$d=\norm{\sigma}$, thus we can now write
\begin{equation}\label{e.diversity'}\tag{\ref{e.diversity}'}
   \div
   =\sum_{i=0}^{\k} \hat{\s}_i\,\left(d^2\varrho\right)^i
\end{equation}
which emphasizes the intuitively obvious fact, that scaling of $\varrho$
(resp. $\rho$) behaves reciprocal to scaling of the distances. Moreover, we
see that the diversity scales (term wise) with (an appropriate power of) the
metric $d$, which means in particular that the task of maximizing the
diversity behaves in its higher order terms (especially the diversity
product) like a constraint on the packing problem determined by the
diversity sum, contrasting the impression one might have gotten
by considering only the Chernov bound \eqref{e.chernov}, which seems
dominated by its highest order term. Consequently we have to 
control $\p$ constrained on the unit sphere $S_{d}$.
In summary the homogeneity property 
(\ref{e.diversity'}) scales all orders of
diversity by the pairwise metric distances, turning the diversity orders
$\ge 2$ into local quantities.  
Thus maximizing diversity corresponds roughly to locally
maximizing the diversity product while globally maximizing the diversity
sum (constrained packing problem).
The behavior of the diversity product on large scales becomes unimportant
due to the contributions of the lower order terms.
Let us therefore perform a Lagrangian analysis for the diversity product
constrained on the unit sphere.

\begin{figure}[htb]
   \center{\includegraphics[height=4cm,width=5cm]{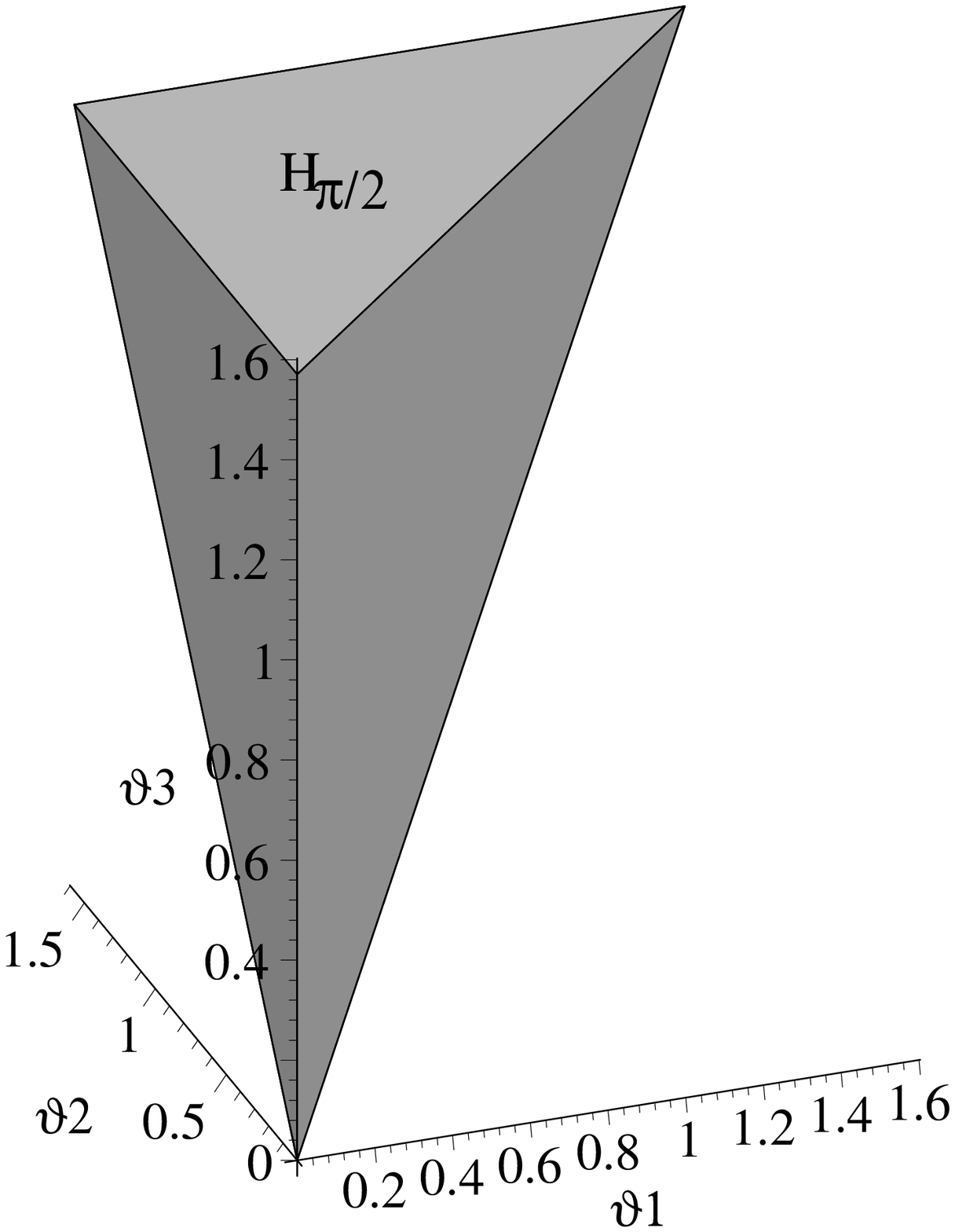}}
   \caption{\label{fig.simplex}$\bar\Theta$ for $\k=3$}
\end{figure}

{\bf Lagrangian analysis:} 
The non-coherent channel diversity sum and product and the corresponding lower
bounds for their coherent channel analogues (by embedding), are
functions of type $H(\vartheta)=\sum_{i=1}^{\k} h(\vartheta_i)$ or
$K(\vartheta)=\prod_{i=1}^{\k} h(\vartheta_i)$ with either $h=\sin^2$ 
(for $\duc^2$ (\ref{e.duc}), $\puc^2$ (\ref{e.p})) or 
$h=2(1-\cos)$ (for $\ddist^2$ (\ref{e.dist}), $\pdist^2$ (\ref{e.pdist})). 
Their domain of definition is the closed simplex $\obar{\Theta}$ of
principal angles (see figure \ref{fig.simplex})
\begin{equation}
  \obar{\Theta}\mdef
    \left\{\left.
      \vartheta=(\vartheta_1,\dots,\vartheta_{\k}) \:\right|\:
      0\le\vartheta_1\le\cdots\le\vartheta_{\k}\le\frac{\pi}{2}
    \right\}
\end{equation}
(the open simplex being 
$\Theta\mdef\{\vartheta\in\RR^{\k}\,|\,
 0<\vartheta_1<\cdots<\vartheta_{\k}<\pi/2\}$) 
but since the principal angles vary like the identity map
$\id$ for $0\le\vartheta_i\le\pi/2$ but extend to 
$\pi-\id$ for $\pi/2<\vartheta_i<\pi$ (considered as a function on the
aperture angle), the function $h$ (i.e. $1-\cos$) fails in general to be
differentiable transversal to the closed 
facet $H_{\pi/2}$ of $\obar{\Theta}$ containing
$\vartheta_{\k}=\pi/2$. Transversal to 
any other edge $h$ is smooth, of course. 
In order to apply the classical Lagrangian formalism of constrained
optimization problems in $\RR^{\k}$ to the present situation, we must
convince ourselves, that the non-smooth edges and the $\pi/2$-facet of
$\obar{\Theta}$ do not interfere. Our next task therefore consists of an
appropriate decomposition of $\obar{\Theta}$ into smooth pieces, which
decompose the optimization into a series of smaller tasks of one single
type, solvable simultaneously in $\RR^{l}$, $l\le\k$ (formula
(\ref{e.extremal-points}) shows the resulting problem formulation).

We need a little bit more notation. 
Let $\obar{\Theta}_< \mdef \obar{\Theta} \setminus H_{\pi/2}$, 
and $\d\Theta_< \mdef \obar{\Theta}_< \setminus \Theta$ the $C^0$ boundary
manifold of $\Theta$ with the problematic facet removed. For
$l=0,\dots,k$ the faces contained in $\obar{\Theta}_<$ of dimension $l$ are
given by  
\begin{equation*}\begin{split}
 & \d^{(l)}\Theta_< \mdef\\
 &    \left\{\left.
        \begin{aligned}
             0 & =\vartheta_1 = \dots =\vartheta_{p_1} \\
         & < \vartheta_{p_1+1} = \dots =\vartheta_{p_2} \\
         & \quad\vdots \\
         & < \vartheta_{p_l+1} = \dots =\vartheta_{\k} 
        \end{aligned}
      \,\right|\,
         \begin{gathered}
            p_1<\dots<p_l \\
            p_i = i-1\dots\k-(l-i+1)
         \end{gathered}
      \right\}
\end{split}\end{equation*}
thus
$\d^{(l)}\Theta_<$ consists exactly of those faces in $\obar{\Theta}_<$,
which are given by $p_1$ (possibly $=0$) zero angles followed by $l$
'blocks' each of equal nonzero angles, in increasing order,   
in particular $\d^{(0)}\Theta_<=\{0\}$, $\d^{(\k)}\Theta_<=\Theta$. 
Each face $\d^{(l)}\Theta_<$ is a smooth submanifold of $\d\Theta_<$, with
$\d(\d^{(l)}\Theta_<)=\d^{(l-1)}\Theta_<$
and 
$\obar{\Theta}_<=\dot{\cup}_{l=1}^{\k} \d^{(l)}\Theta_<$.
The tangent spaces are 
\begin{equation*}\begin{split}
 & T\left( \d^{(l)}\Theta_<  \right) = \\
 &     \left\{\!
          \sum_{i=1}^l \lambda_i (e_{p_i+1}+\dots+e_{p_{i+1}})
          \left|
            \begin{gathered}
               \lambda_i \in \RR \\ 
               p_1<\dots<p_l<p_{l+1}=\k \\
               p_i = i-1\dots\k-(l-i+1)
            \end{gathered}            
          \right.\!\!
       \right\}
\end{split}\end{equation*}
Then we have
\begin{lemma}\textin
   Given $H(\vartheta)=\sum_{i=1}^{\k} h(\vartheta_i)$, 
   $K(\vartheta)=\prod_{i=1}^{\k} h(\vartheta_i)$
   on $\obar{\Theta}_<$ with $h\in C^{\infty}([0,\pi/2[)$ (this means, that
   $h$ is differentiable from the right in $0$) and $h'(0)=0$. Then
   for $l=0,\dots,\k$ 
   \begin{gather}
      \grad H_{|\d^{(l)}\Theta_<} \in T\left( \d^{(l)}\Theta_<  \right) \\
      \grad K_{|\d^{(l)}\Theta_<} \in T\left( \d^{(l)}\Theta_<  \right) 
   \end{gather}\vspace{-4ex}
\end{lemma}
thus restricting the gradients remains intrinsic.
\\[1ex]
\begin{proof}\myline
   We have $\grad H(\vartheta)=(h'(\vartheta_1),\dots,h'(\vartheta_{\k}))$.
   Since $h'(0)=0$ the symmetry of $H$ ensures\myline
   $\grad H_{|\d^{(l)}\Theta_<}(\vartheta)
    =\sum_{i=1}^l h'(\vartheta_{p_i+1})(e_{p_i+1}+\dots+e_{p_{i+1}})
    \in T\left( \d^{(l)}\Theta_<\right)$,
   with $\lambda_i=h'(\vartheta_{p_i+1})$.
   Similarly, 
   $\grad K(\vartheta)=(h'(\vartheta_i)\prod_{j\neq i}h(\vartheta_j))_i$,
   thus $\grad K_{|\d^{(l)}\Theta_<}(\vartheta)=0$ (for $p_1>0$) or 
   $\grad K_{|\d^{(l)}\Theta_<}(\vartheta)
   =\sum_{i=1}^l \lambda_i (e_{p_i+1}+\dots+e_{p_{i+1}})
   \in T\left( \d^{(l)}\Theta_<\right)$ 
   with \myline
   $\lambda_i
    =h'(\vartheta_{p_i+1})
     h(\vartheta_{p_i+1})^{p_{i+1}-p_i-1}
     \prod_{j\neq i}h(\vartheta_{p_j+1})^{p_{j+1}-p_j}$.
\end{proof}

The lemma ensures, that the Lagrangian functional $F=f-\lambda (g-\delta)$
on a neighborhood of $\obar{\Theta}$
for critical points of $f$ obeying the constraint $g=\delta$ 
($f,g$ either $H$ or $K$, $\delta\in\RR$) applies to the boundary
$\d\Theta_<$. Since  
$f$ and $g$ are not necessarily differentiable transversal to $H_{\pi/2}$,
extremal points in $\obar{\Theta}$ lie in the set
$\{g=\delta\} \cap H_{\pi/2} 
  \:\cup\:
  \obar{\Theta}_< \cap \{DF=0\}
$
(recall, that $\cap$ has higher precedence than $\cup$ and $D$ denotes
differentiation).
For a more unified treatment we define 
$\obar{\Theta}^{(l)}
 \mdef\{0\le\vartheta_1\le\cdots\le\vartheta_l\le\pi/2\}$
(in particular $\obar{\Theta}^{(0)}=\emptyset$ and 
 $\obar{\Theta}^{(\k)}=\obar{\Theta}$)
and recall, that on the one hand $f$, $g$ are
differentiable tangential to $H_{\pi/2}$ and on the other hand 
$\obar{\Theta}^{l-1} \cup \{\vartheta_l=\dots=\vartheta_{\k}=\pi/2\}$,
$l=1,\dots,\k-1$ exhausts $H_{\pi/2}$. 
This leads to the following recursion scheme: For
$l=0,\dots,\k$ set 
\begin{equation}
   \begin{aligned}
      & H^{(l)}(\vartheta_1,\dots,\vartheta_l)
          = \sum_{i=1}^l h(\vartheta_i) + h(\tfrac{\pi}{2})(\k-l) \\
      & K^{(l)}(\vartheta_1,\dots,\vartheta_l)
          =\prod_{i=1}^l h(\vartheta_i) \cdot h(\tfrac{\pi}{2})^{(\k-l)}
   \end{aligned}
\end{equation}
and $f^{(l)}, g^{(l)}$ given by either $H^{(l)}$ or $K^{(l)}$.
Then the extremal points in $\obar{\Theta}$ lie in the set
\begin{equation}\label{e.extremal-points}
  \bigcup_{l=0}^{\k} 
    \left( \{DF^{(l)}=0\} \cap 
            \obar{\Theta}^{(l)}_< \cup
            \{\vartheta_{l+1}=\dots=\vartheta_{\k}=\frac{\pi}{2}\}
    \right)
\end{equation}
whereas $F^{(l)}\mdef f^{(l)}-\lambda (g^{(l)}-\delta)$.
Furthermore, for $h$ monotonically increasing on $[0,\pi/2]$ and zero at
$0$, the conditions $g^{(l)}=\delta$ forces 
$\delta\in[0,h(\tfrac{\pi}{2})\k]$
(resp. $\delta\in[0,h(\tfrac{\pi}{2})^\k]$) and in the 
first case $l\ge \lceil \k-\delta/h(\tfrac{\pi}{2})\rceil \mfed \ubar{l}$,
thus $l$ restricts to $\{\ubar{l},\dots,\k\}$.

Let us now start the Lagrangian analysis of the diversity (resp. with the
analysis of the various diversity sums and products). The non-coherent channel
diversity 
sum/product $(\duc)(\puc)$ as well as the lower bounds 
$(\ddist)(\pdist)$ for the
coherent channel analogues depend on the $U(\k)$-fibers only (moreover they
depend only on the principal angles). By left invariance we can always
assume $\Psi=\kbein$ and consider 
the diversity terms as functions (marked with an ${}_{\ubar{0}}$) on the
single argument $\Phi$, for which $\generate{\Phi}$ is separated by
principal angles $\vartheta \in \obar{\Theta}$ from $\generate{\kbein}$.
\\[1ex]
{\bf $\boldsymbol{{\puc}_{\ubar{0}\big|S_{\duc^2}(\delta)}^2}$:}\myline
In order to find the maximum of \puc in the unit distance sphere
we constrain $f={\puc}_{\ubar{0}}^2=\prod_i\sin^2$ on $S_{\duc^2}(\delta)$,
by setting $g={\duc}_{\ubar{0}}^2=\sum_i\sin^2$. 
Thus we get the Lagrangian functional 
$F(\vartheta)
 =\prod_i\sin^2\vartheta_i-\lambda(\sum_i\sin^2\vartheta_i-\delta)$,
$0<\delta\le\k$. Here we have $\ubar{l}=\lceil\k-\delta\rceil$ and from
\eqref{e.extremal-points} we get for each
$l\in\{\ubar{l},\dots,\k\}$ 
\begin{equation*}\begin{split}
 & \{DF^{(l)}=0\}\cap\obar{\Theta}^{(l)}_<
     = \left\{g^{(l)}=\delta\right\}\,\cap\\
 &     \left\{\left.
         \begin{aligned}
             0 & = \vartheta_1=\dots=\vartheta_{p_1}\\
               & < \vartheta_{p_1+1}\le\dots\le\vartheta_l<\frac{\pi}{2}
         \end{aligned}
         \,\right|
         \lambda = \prod_{i\neq j}^l \sin^2\vartheta_i
         ,\,\forall_{p_1<j\le l}
       \right\}
\end{split}\end{equation*}
From this we get extremal points with $f\neq 0$ only for $p_1=0$, 
$\vartheta_1=\dots=\vartheta_l=\theta^{(l)}$ with 
$l\sin^2\theta^{(l)} = \delta-(\k-l)$ and
$f^{(l)}=(\sin^2\theta^{(l)})^l=(1-\tfrac{\k-\delta}{l})^l$, monotonically
increasing with $l$, therefore
\begin{equation}\label{e.max-pucduc}
   \max {\puc}_{\ubar{0}\big|S_{\duc^2}(\delta)}^2 
      = f^{(\k)}
      = \left(\frac{\delta}{\k}\right)^{\k}
\end{equation}
attained in 
$\vartheta_1=\dots\vartheta_{\k}=\theta$,
$\sin^2\theta = \delta/\k$.
\begin{conclusion}\label{conc.cg-points}\textin
   Locally the code points of ${\cal C}^G$ for the non-coherent channel have
   to be distributed with as many of their 
   pairwise principal angles to be nonzero and equal in modulus as possible.
\end{conclusion}

In principle, the same holds for the coherent channel, if we consider \pkc and
\dkc as functions of $\sigmakc$: The maximum diversity product is attained
for $\sigmakc_1=\dots=\sigmakc_{\k}=\mathfrak{s}$, $\mathfrak{s}^2=\delta/\k$,
$\pkc^2=\mathfrak{s}^{2\k}=(\delta/\k)^{\k}$, but it seems difficult to
characterize all $\Phi\in S_{\dkc^2}(\delta)$ subject to 
$\sigmakc_1=\dots\sigmakc_{\k}=\mathfrak{s}$.
Let us embed ${\cal C}^G$ into \vkn instead and investigate the question,
which conditions have to be imposed on ${\cal C}^G$ in order to achieve some
diversity gain in terms of $\ddist$ and \pdist. 
\\[1ex]
{\bf $\boldsymbol{\ddist^2_{\ubar{0}\big|S_{\duc^2}(\delta)}}$:}\myline
For the metric fiber distance $\ddist$ constrained on $S_{\duc^2}(\delta)$
the Lagrangian functional reads 
$F(\vartheta)=\ddist^2_{\ubar{0}}(\vartheta) 
             -\lambda ({\duc}_{\ubar{0}}^2(\vartheta)-\delta)
             =2k-2\sum_i\cos\vartheta_i
             -\lambda(\sum_i\sin^2\vartheta_i-\delta)$.
Again we have $0<\delta\le\k$, $\ubar{l}=\lceil\k-\delta\rceil$ and from
\eqref{e.extremal-points} we now get for each
$l\in\{\ubar{l},\dots,\k\}$ 
\begin{equation*}\begin{split}
 & \{DF^{(l)}=0\}\cap\obar{\Theta}^{(l)}_<
     = \left\{g^{(l)}=\delta\right\}\,\cap\\
 &     \left\{\left.
         \begin{aligned}
             0 & = \vartheta_1=\dots=\vartheta_{p_1}\\
               & < \vartheta_{p_1+1}\le\dots\le\vartheta_l<\frac{\pi}{2}
         \end{aligned}
         \,\right|
         \lambda = \frac{1}{\cos\vartheta_j}
         \mathcomma\forall_{p_1<j\le l}
       \right\}
\end{split}\end{equation*}
Extremal points are contained in 
$0=\vartheta_1=\dots=\vartheta_{p_1}$, 
$\vartheta_{p_1+1}=\dots=\vartheta_l=\theta^{(l)}$ subject to
$(l-p_1)\sin^2\theta^{(l)}=\delta-(\k-l)$ for 
$l=\ubar{l}=\lceil\k-\delta\rceil,\dots,\k$ and 
$p_1=0,\dots,\obar{p}_1=\lfloor\k-\delta\rfloor$, with
$f^{(l)}_{p_1}=2(l-p_1)(1-\cos\theta^{(l)})+2(\k-l)
 =2(l-p_1)(1-\sqrt{1-q})+2(\k-l)$, 
$q\mdef\tfrac{\delta-\k+l}{l-p_1}=\sin^2\theta^{(l)}$. 
For fixed $l$ the function $\mapto{p_1}{f^{(l)}}_{p_1}$ is monotonically
increasing (by analyzing the derivative, where defined) and we find
$\ubar{f}^{(l)}\mdef\min_{p_1}f^{(l)}_{p_1}=f^{(l)}_0$ and 
$\obar{f}^{(l)}\mdef\max_{p_1}f^{(l)}_{p_1}=f^{(l)}_{\obar{p_1}}$. As
functions of $l$ both terms turn out to be monotonically decreasing 
(by analyzing the derivatives with respect to $l$) and we find
\begin{subequations}\label{e.maxmin-distduc}
\begin{gather}\label{e.min-distduc}
       \min \ddist^2_{\ubar{0}\big|S_{\duc^2}(\delta)} 
          = \ubar{f}^{(\k)}
          =  2\k\left(1-\sqrt{1-\delta/\k}\right) 
       \intertext{attained in $\vartheta_1=\dots\vartheta_{\k}=\theta$,
                              $\sin^2\theta = \delta/\k$, and}
       \label{e.max-distduc}
       \max \ddist^2_{\ubar{0}\big|S_{\duc^2}(\delta)} 
          = \obar{f}^{(\ubar{l})}
          =  2\left(1-\sqrt{1-(\delta-\lfloor\delta\rfloor)}\right) 
            +2 \lfloor\delta\rfloor
\end{gather}
attained for $\vartheta_1=\dots\vartheta_{\lfloor\k-\delta\rfloor}=0$,
             $\sin^2\vartheta_{\lceil\k-\delta\rceil} =
              \delta-\lfloor\delta\rfloor$,
             $\vartheta_{\lceil\k-\delta\rceil+1}=\dots\vartheta_{\k}=\pi/2$.
\end{subequations}
\\[2ex]
{\bf $\boldsymbol{\pdist^2_{\ubar{0}\big|S_{\duc^2}(\delta)}}$:}\myline
Examining \pdist instead of $\ddist$ we have the Lagrangian
$F(\vartheta)=2^k\prod_i(1-\cos\vartheta_i)
             -\lambda(\sum_i\sin^2\vartheta_i-\delta)$
and 
\begin{equation*}\begin{split}
 & \{DF^{(l)}=0\}\cap\obar{\Theta}^{(l)}_<
     = \left\{g^{(l)}=\delta\right\}\,\cap\\
 &     \left\{\left.
         \begin{aligned}
             0 & = \vartheta_1=\dots=\vartheta_{p_1}\\
               & < \vartheta_{p_1+1}\le\dots\le\vartheta_l<\frac{\pi}{2}
         \end{aligned}
         \right|
         \begin{aligned}
            \lambda = \frac{2^{\k-1}}{\cos\vartheta_j}
                      \prod_{i\neq j}^l (1-\cos\vartheta_i) & \\
            \forall_{p_1<j\le l} & 
         \end{aligned}
       \right\}
\end{split}\end{equation*}
From this we get extremal points with $f\neq 0$ only for $p_1=0$,  
$\vartheta_1=\dots=\vartheta_l=\theta^{(l)}$ with  
$l\sin^2\theta^{(l)} = \delta-(\k-l)$ and 
$f^{(l)}=2^{\k}(1-\cos\theta^{(l)})^l
 =2^{\k}(1-\sqrt{\tfrac{\k-\delta}{l}})^l$, monotonically increasing with
$l$ and therfore
\begin{equation}\label{e.max-pdistduc}
   \max \pdist^2_{\ubar{0}\big|S_{\duc^2}(\delta)} 
      = f^{(\k)}
      = 2^{\k}\left( 1-\sqrt{1-\delta/\k} \right)^{\k}
\end{equation}
in $\vartheta_1=\dots\vartheta_{\k}=\theta$,
   $\sin^2\theta = \delta/\k$.
\begin{remark}\textin
   If we had constrained \pdist on $S_{\ddist^2}(\delta)$, $0<\delta\le2\k$
   we would have gotten  
   \begin{equation}\label{e.max-pdistdist}
      \max \pdist^2_{\ubar{0}\big|S_{\ddist^2}(\delta)} 
      = \left(\frac{\delta}{\k}\right)^{\k}
   \end{equation}
   attained in
   $\vartheta_1=\dots\vartheta_{\k}=\theta$,
   $2(1-\cos\theta) = \delta/\k$,
   by analogy to the case of \puc constrained on \duc. 
   The apparent discrepancy of \eqref{e.max-pdistduc} to
   \eqref{e.max-pdistdist} is caused by \eqref{e.max-distduc}, since at 
   $\vartheta_1=\dots=\vartheta_{\k}=\theta$ 
   $\ddist^2_{\ubar{0}}$ attains its minimum 
   $\tilde{\delta}=2\k(1-\sqrt{1-\delta/\k})$ on $S_{\duc^2}(\delta)$ which
   yields (with \eqref{e.max-pdistdist})
   $\pdist^2_{\ubar{0}|\ddist^2_{\ubar{0}}=\tilde{\delta}}
   =(\tilde{\delta}/\k)^{\k}=2^{\k}(1-\sqrt{1-\delta/\k})^{\k}$, which
   coincides with \eqref{e.max-pdistduc}.
\end{remark} 
\begin{conclusion}\label{conc.cg-points-in-v}\textin
   What remains is the general rule,
   that for $\varrhokc = \tfrac{\rho\n}{4\k}\gg1$ (compare (\ref{e.varrhokc})) in
   favor of \pdist instead of \ddist one should distribute the pairwise
   principal angles in ${\cal C}^G$ locally to be all nonzero and equal
   in modulus (by \eqref{e.max-pdistduc}). This coincides with the
   preferred strategy \eqref{e.max-pucduc} for the non-coherent channel code. 
   For $\varrhokc\le 1$, when the higher order
   diversity terms become less important, it might be better to
   distribute the pairwise principal angles globally to maximize \ddist, thus
   separating points in ${\cal C}^G$ by as many of the principal angles to be
   zero such that the remaining ones attain large values.
\end{conclusion}
{\bf $\boldsymbol{\pdist^2_{\ubar{0}\big|S_{\puc^2}(\delta)}}$:}\myline
Finally, to get an product analogue of \eqref{e.maxmin-distduc} 
we analyze
$\pdist^2_{\ubar{0}}$ constrained on $S_{\puc^2}(\delta)$, $0<\delta\le 1$,
thus the Lagrangian reads
$F(\vartheta)=2^k\prod_i(1-\cos\vartheta_i)
             -\lambda(\prod_i\sin^2\vartheta_i-\delta)$ 
and for $l\in\{0,\dots,\k\}$ we have
\begin{equation*}\begin{split}
 & \{DF^{(l)}=0\}\cap\obar{\Theta}^{(l)}_<
     = \left\{g^{(l)}=\delta\right\}\,\cap\\
 &     \left\{\left.
         \begin{aligned}
             0 & = \vartheta_1=\dots=\vartheta_{p_1}\\
               & < \vartheta_{p_1+1}\le\dots\le\vartheta_l<\frac{\pi}{2}
         \end{aligned}
         \right|
         \begin{aligned}
         \lambda = \frac{2^{\k-1}}{\cos\vartheta_j}
                   \prod_{i\neq j}^l 
                      \frac{1-\cos\vartheta_i}{\sin^2\vartheta_i} &\\
         \forall_{p_1<j\le l} &
         \end{aligned}
       \right\}
\end{split}\end{equation*}
$\delta>0$ forces $p_1=0$ and we get 
$\vartheta_1=\dots=\vartheta_l=\theta^{(l)}$,
$\sin^{2l}\theta^{(l)}=\delta$, thus
$f^{(l)}=2^{\k}(1-\cos\theta^{(l)})^l
 =2^{\k}(1-\sqrt{1-\delta^{1/l}})^l$ which is monotonically decreasing
in $l$, so
\begin{subequations}\label{e.maxmin-pdistpuc}
   \begin{gather}
       \min \pdist^2_{\ubar{0}\big|S_{\puc^2}(\delta)} 
          = f^{(\k)}
          =  2^{\k}\left( 1-\sqrt{1-\delta^{1/\k}} \right)^{\k}
     \intertext{attained in $\vartheta_1=\dots\vartheta_{\k}=\theta$,
                            $\sin^2\theta = \delta^{1/\k}$, and}
       \max \pdist^2_{\ubar{0}\big|S_{\puc^2}(\delta)} 
          = f^{(1)}
          =  2^{\k}\left( 1-\sqrt{1-\delta} \right) 
   \end{gather}
   attained in $\sin^2\vartheta_1 = \delta$,
               $\vartheta_2=\dots=\vartheta_{\k}=\pi/2$
\end{subequations}
(whereas $f^{(0)}$ is contained in $f^{(1)}$ when $\delta=1$).

Observing, that the minima in \eqref{e.maxmin-distduc} and
\eqref{e.maxmin-pdistpuc} are monotonically increasing in $\delta$ we end
up with 
\begin{proposition}\label{prop.explicit-embedding-bounds}\textin
   In the situation of Proposition \ref{prop.code-embedding} we
   have
   \begin{align}\label{e.distmin-ducmin}
     && \ddist^{2\min} 
       & \ge \min\ddist^2_{\ubar{0}\big|S_{\duc^2}(\duc^{2\min})}\\ \notag
     &&&    = 2\k \left(1-\sqrt{1-\duc^{2\min}/\k}\right)
             \ge \duc^{2\min} \\
     \label{e.pdistmin-pucmin}
     && \pdist^{2\min} 
       & \ge \min \pdist^2_{\ubar{0}\big|S_{\puc^2}(\puc^{2\min})}\\ \notag
     &&&    = 2^{\k} \left( 1-\sqrt{1-{\puc^{2\min}}^{1/\k}} \right)^{\k}
             \ge \puc^{2\min}
   \end{align}\vspace{-4ex}
\end{proposition}
The benefit of this proposition is, that it relates $\ddist^{\min}$
(resp. $\pdist^{\min}$) directly to $\duc^{\min}$ (resp. $\puc^{\min}$),
regardless if the minimum distances (resp. diversity products) are realized
by the same pair of points or not. 
\\[1ex]
\begin{proof}\myline
   Its only left up to show the second inequality in each formula, which is
   elementary (setting $x\mdef \duc^{2\min}$ or $x\mdef \puc^{2\min}$,
   respectively):\myline
   \eqref{e.distmin-ducmin}
   $\equivalentshort
    \sqrt{1-x/\k}\le 1-x/2\k
    \equivalentshort
    1-x/\k\le (1-x/2\k)^2
           = 1-x/\k + x^2/4\k^2$.\myline
   \eqref{e.pdistmin-pucmin} 
   $\equivalentshort
    \sqrt{1-x^{1/\k}}\le 1-x^{1/k}/2
    \equivalentshort
    1-x^{1/\k}\le (1-x^{1/k}/2)^2
                       = 1-x^{1/\k} + x^{2/\k}/4$
\end{proof}
\section{Conclusions}
This work should be seen as a second step towards a geometry based analysis
of general space time block codes, inspired by the results in \cite{hen-transinf1},
opening the door to potentially high performing space time
block codes, when $\n\gg\k$.
The various estimates and interrelations explored in this work assemble the
following overall picture: 
\begin{itemize}
\item {\bf Diversity monotony:}
   The performance analysis revealed nice embedding properties (with respect
   to $\gkn\subset\vkn$) of the
   diversity quantities  
   (Corollaries \ref{cor.s-embedding}, \ref{cor.min-s-embedding},
   (\ref{e.varrhouc-estimate})), leading to a diversity growth
   (Proposition \ref{prop.code-embedding}) in the
   transition from the non-coherent channel to the coherent channel.
   This turned out to be due to the various
   invariance properties satisfied by the diversity, though
   tied to distinct underlying topologies of 
   the coding spaces induced by the maximum likelihood receiver metrics.
   Moreover, for the diversity sum and product, more explicit estimates
   have been derived (Proposition \ref{prop.explicit-embedding-bounds}).
\item {\bf Complexity reduction:}
   Embeddings of both \gkn and \Uk into \vkn can be used to construct codes
   on \vkn from 'smaller' pieces 
   (Theorem \ref{thm.code-diversity-compose}), both of them being
   already in the focus of current research. 
   The other way round, given an non-coherent channel space time code and a 'small'
   coherent channel code, the performance of the resulting (larger dimensional)
   product code on \vkn 
   is lower bounded by the diversity expressions stated in the theorem. Thus
   the design complexity has been reduced to the smaller problems on \gkn and
   \Uk. Together with Proposition \ref{prop.dmin-lowerbound}
   this opens the door to potentially high performing space time
   block codes, when $\n\gg\k$. 
   As already indicated in the introduction this may be of some importance
   in the context of space frequency codes also. 
\item {\bf Localization:}
   The local nature of the higher order diversity quantities turns space
   time coding into a constrained packing problem. 
   The diversity sum still represents a major criteria, locally superposed
   by the diversity product as a rigidity constraint: The optimal code in
   the high SNR regime is packed as 'diagonal' as possible, uniformly
   maximizing the principal 
   angles (Conclusion \ref{conc.cg-points} and \ref{conc.cg-points-in-v}).
\end{itemize}
There are still many open issues. Some immediate will be listed next.
The explicit bounds of
Proposition \ref{prop.explicit-embedding-bounds} are very coarse and
improvements are necessary. 
Moreover, it would be desirable to obtain further decompositions in
Theorem \ref{thm.code-diversity-compose}. 
Furthermore this work has to be related to the differential coding scheme
\cite{hoc.swe}, which benefits from high rates compared to non
differential codes.
Finally it remains the challenge of effective high dimensional code
construction (especially for the non-coherent channel) with low complexity
decoding properties.

\section*{Acknowledgment}
I would like to thank Eduard Jorswieck, Peter Jung, and \mbox{Aydin} Sezgin for
reading the manuscript and helpful comments.
%
%
%
%
%
%
%
%
%
%
%
%

%
%
%
%
%
\end{document}